# Epitaxial single-crystal growth of transition metal dichalcogenide monolayers *via* atomic sawtooth Au surface

Soo Ho Choi, Hyung-Jin Kim, Bumsub Song, Yong In Kim, Gyeongtak Han, Hayoung Ko, Stephen Boandoh, Ji Hoon Choi, Chang Seok Oh, Hyun Je Cho, Jeong Won Jin, Seok Joon Yun, Bong Gyu Shin, Hu Young Jeong, Young-Min Kim, Young-Kyu Han*, Young Hee Lee*, Soo Min Kim*, and Ki Kang Kim*

Dr. S. H. Choi, Mr. B. Song, Dr. S. Boandoh, Dr. S. J. Yun, Prof. Y. -M. Kim, Prof. Y. H. Lee, Prof. K. K. Kim
Center for Integrated Nanostructure Physics (CINAP), Institute for Basic Science (IBS), Sungkyunkwan University, Suwon, 16419, Republic of Korea

Dr. H. -J. Kim, Prof. Y. -K. Han
Department of Energy and Materials Engineering, Dongguk University, Seoul, 04620, Republic of Korea

Mr. B. Song, Mr. Y. I. Kim, Mr. G. Han, Mr. H. Ko, Mr. J. H. Choi, Mr. C. S. Oh, Mr. H. J. Cho, Ms. J. W. Jin, Prof. Y. -M. Kim, Prof. Y. H. Lee, Prof. K. K. Kim
Department of Energy Science, Sungkyunkwan University, Suwon, 16419, Republic of Korea

Dr. B. G. Shin
Max-Planck Institute for Solid State Research, Heisenbergstrasse 1, 70569 Stuttgart, Germany

Prof. H. Y. Jeong
UNIST Central Research Facilities, School of Materials Science and Engineering, UNIST, Ulsan, 44919, Republic of Korea

Prof. S. M. Kim
Department of Chemistry, Sookmyung Women's University, Seoul, 14072, Republic of Korea

E-mail: ykenergy@dongguk.edu, leeyoung@skku.edu, soominkim@sookmyung.ac.kr, kikangkim@skku.edu

Growth of two-dimensional van der Waals layered single-crystal (SC) films is highly desired to manifest intrinsic material sciences and unprecedented devices for industrial applications. While wafer-scale SC hexagonal boron nitride film has been successfully grown, an ideal growth platform for diatomic transition metal dichalcogenide (TMdC) film has not been established to date. Here, we report the SC growth of TMdC monolayers in a centimeter scale via atomic sawtooth gold surface as a universal growth template. Atomic tooth-gullet surface is constructed by the one-step solidification of liquid gold, evidenced by transmission-electron-microscopy. Anisotropic adsorption energy of TMdC cluster, confirmed by density-functional calculations, prevails at the periodic atomic-step edge to yield unidirectional epitaxial growth of triangular TMdC grains, eventually forming the SC film, regardless of Miller indices. Growth using atomic sawtooth gold surface as a universal growth template is demonstrated for several TMdC monolayer films, including $WS_2$, $WSe_2$, $MoS_2$, $MoSe_2/WSe_2$ heterostructure, and $W_{1-x}Mo_xS_2$ alloy. Our strategy provides a general avenue for the SC growth of diatomic van der Waals heterostructures in a wafer scale, to further facilitate the applications of TMdCs in post silicon technology.

Two-dimensional (2D) van der Waals (vdW) layered materials, including graphene, transition metal dichalcogenide (TMdC), and hexagonal boron nitride (hBN), have been extensively investigated in various fields owing to their unusual physical and chemical properties for developing unprecedented electronics, valleytronics, twistronics, spintronics, and catalysts.[1-7] For example, an atomically thin vertical transistor on graphene/hBN/$MoS_2$ heterostructure, PN diode with n-$MoS_2$/p-$WSe_2$ heterostructure, and coulomb drag transistors in $MoS_2$/graphene heterostructure and multilayer $WSe_2$ have been fabricated within a few micrometers in size.[8-10]

Large-area 2D films have been successfully synthesized by merging randomly oriented 2D grains, inevitably forming polycrystalline films.[11-14] While the intrinsic physical properties of polycrystalline films have been comprehensively investigated in terms of micron-scale grains, its inhomogeneity in a wafer scale limits application to integration challenges associated with structural defects such as grain boundaries and disorder.[15] The growth of single-crystal (SC) monolayer and multilayer graphene films on SC Cu (111) and Cu-Si (111) surfaces has been demonstrated in a wafer scale,[16,17] whereas the growth of diatomic 2D SC film such as hBN and TMdCs remains still complicated due to their non-centrosymmetric structures.[18] Recently, the self-collimation of hBN grains on liquid Au substrate has been proposed for growing diatomic hBN SC film in a wafer scale but further study is required to grow other 2D vdW materials.[19] Moreover, step-edge guided epitaxial growth has been observed in hBN films on SC Cu (111) and (110).[18, 20]

Our primary concern is to discover a growth platform for single-crystal TMdCs in a centimeter scale. While SC hBN and Au (111) vicinal surface have been produced for the growth of SC TMdC film,[19,21] evidence of seamless stitching, other Miller indices, and more importantly, a growth platform for general diatomic TMdCs still require investigation. We report a growth platform for SC TMdCs in a centimeter scale via an atomic sawtooth Au surface, which is evidenced by cross-sectional high-resolution transmission-electron-microscopy (HRTEM) to further verify the grain boundary (GB)-free SC TMdC film, regardless of Miller indices.

The key to achieving SC TMdC growth is to construct the atomic sawtooth Au surface, to serve as a universal growth template, consisting of periodic tooth-gullet step edges to accommodate the self-assembly of unidirectional TMdC grains (**Figure 1**a). The atomic sawtooth Au surface with periodic step edges and low Miller index terraces, including (100), (110), and (111) facets, was prepared through the one-step solidification of liquid Au on W substrate (see Experimental Section). The atomic sawtooth Au (533) surface with the topmost $WS_2$ monolayer is clearly distinguishable through cross-sectional annular dark-field scanning transmission electron microscopy (ADF-STEM) (Figure 1b). A surface with an irregular atomic sawtooth periodicity still yields a robust $WS_2$ monolayer (Supporting Information, Figure S1).

The chemical inertness of Au and negligible solubility of metal and chalcogen atoms in Au film make the epitaxial growth of monolayer TMdC film possible.[22] In contrast, conventional Cu film is highly reactive with chalcogen atoms.[23]

The electron backscatter diffraction (EBSD) inverse pole figure (IPF) maps of the solidified Au foil demonstrate the formation of diverse atomic sawtooth surfaces such as (5,7,11), (169) and (114) on respective (111), (110), and (100) terrace facets (Figures 1c, e, g). The EBSD mappings of high Miller indices are uniformly colored. We note that high-index surfaces were randomly produced after the solidification of liquid Au (Supporting Information, Figures S2, S3, and Table S1), contrary to previous study in which low-index Au (111) surface was obtained using similar preparation procedure.[21] Monolayer $WS_2$ grains were synthesized at 800 °C by injecting ammonium sulfide with a bubbler system onto the W precursor-coated atomic sawtooth Au substrate (see Experimental Section). After the initial growth stage of 5 min, $WS_2$ grains are oriented along a specific direction with coherent triangles in a centimeter scale (Figures 1d, f, h, and Supporting Information, Figure S4), which is persistent with the presence of scarce coherent triangular voids near the final stage of film growth (Supporting Information, Figure S5). These results are in stark contrast with the appearance of triangular and inverted triangular TMdC grains on c-plane sapphire substrates reported by other studies.[24-26]

After a growth period of 20 min, a centimeter-scale SC $WS_2$ film is eventually obtained (Figure 1i). Low energy electron diffraction (LEED) pattern clearly indicates a supercell of $WS_2$ hexagonal lattice, Au (111) terrace facet, and atomic sawtooth Au (221) surface (Figure 1j and inset), congruent with the constructed LEED pattern (Supporting Information, Figure S6). The superposition of 25 LEED patterns in $5 \times 5$ mm$^2$ area clearly confirmed the identical honeycomb lattice to assure SC $WS_2$ film (Figures 1k, l and Supporting Information, Figure S7). The perfect honeycomb $WS_2$ lattice was obtained with a lattice constant of 3.22 Å, indicating that the strain is negligible. This is distinct from previous work involving the growth of strained SC hBN on liquid Au substrate.[19]

To verify the existence of GB-free $WS_2$ film, partially merged $WS_2$ grains grown for a period of 5 min were characterized by HRTEM. After the transfer of the $WS_2$ grains onto the TEM grid, the coherent $WS_2$ grains are preserved (**Figure 2**a). ADF-STEM image depicts a hexagonal crystal structure with d-spacings of 0.28 and 0.16 nm, corresponding to the (10-10) and (11-20) planes of 2H-$WS_2$, consistent with previous work (Figure 2b).[22] The stacked six selective-area-electron-diffraction (SAED) patterns (regions I-VI of Figure 2a) demonstrate the exclusive overlap of the hexagonal dots, confirming the coherent lattice orientation (Figure 2c).

The grain boundary in 60° triangular region at the junction between two merged $WS_2$ grains is carefully analyzed (Figure 2d and inset). The monolithic fringes of heavy W atoms are observed without the trace of GB lines (Figure 2e), in contrast with polycrystalline $WS_2$, which possesses structural defects at GBs (Supporting Information, Figure S8). The higher magnification images of three regions in Figure 2e and their stacked fast Fourier transform (FFT) patterns clearly demonstrate the absence of GB lines in the merged regions (Figure 2f). Similar results are also obtained for other merged regions (Supporting Information, Figure S9). The absence of GB lines is analyzed in a macroscopic scale via oxidation in air at 350 °C.[27] No prominent oxidized GB lines are visible between aligned $WS_2$ grains based on SEM and Raman intensity mapping, while oxidized GB lines are apparent between misaligned polycrystalline $WS_2$ grains (Figures 2g-j, and Supporting Information, Figure S10).

Density-functional theory (DFT) calculations were carried out to elucidate the underlying mechanism of the coherent $WS_2$ grain growth on an atomic sawtooth Au surface. We presumed that the growth proceeds with the formation of a $W_3S_6$ cluster, followed by surface migration, nucleation, and growth.[28] A $W_3S_6$ cluster adsorbed onto a flat terrace can freely migrate and anchor at B step edge, owing to preferable energetics with a relatively larger adsorption energy of 1.43 eV at the Au (221) step edge of the Au (111) terrace facet (Supporting Information, Figure S11 and Table S2). The energetics rely on the rotation angle of the $W_3S_6$ cluster (**Figures 3**a, b). Sulfur atoms energetically favor 0° at the step edge over other angles independent of Miller indices (Figure 3c and Supporting Information, Table S3). Another metastable angle is found at 60° with a rotational barrier height of 0.25–0.75 eV, which varies with Miller indices. Such barriers can be overcome at a growth temperature of 800 °C to rotate the $W_3S_6$ cluster from 60° to the energetically favorable 0° angle. This theory is congruent with experimental results, where unidirectional $WS_2$ grains were grown at an elevated growth temperature of 800–880 °C. When the thermal energy is not sufficient to overcome the rotational barrier height, i.e., at low growth temperature such as 750 °C, a proportion of inverted triangular $WS_2$ grains were grown (Supporting Information, Figure S12). In addition, if the terrace facet is consistent over the entire region, the $WS_2$ grains are unidirectionally aligned, regardless of Miller indices. Coherent triangular $WS_2$ grains are consistently obtained, despite the possible generation of several other Miller indices in addition to the main (236) and (2,10,25) surfaces at fixed (111) and (100) terrace facets, respectively (Supporting Information, Figure S11 and Table S2). This is again attributed to the energetically favorable adsorption of $W_3S_6$ clusters on a specific 'B step edge' of an atomic sawtooth surface confirmed by DFT calculations.

Scanning tunneling microscopy (STM) image of the $WS_2$ film grown on, for example, an Au (221) surface, shows aligned hexagonal S atoms along its regular periodic linear features (Figure 3d). Such a periodicity coincides with a terrace width of Au (221), corresponding to the step edge (Supporting Information, Figure S13). STM simulation of a $WS_2$ film on Au (221) substrate, based on the LEED analysis in Figure 1j, demonstrates that the lower height at W sites in the linear feature is attributed to interactions between the step edge of Au (221) and $WS_2$ via charge transfer from Au to the d orbitals of W in $WS_2$ (Figure 3e).[29] A structural model of $WS_2$ monolayer on Au (221) surface including a supercell of $WS_2$ across the step edges for a doubled terrace width in unit cell is constructed (see Experimental Section and Figure 3f). Along the step edges ([01-1]), ×9 of $WS_2$ is commensurate with ×10 of Au (221), indicative of negligible strain formation.

The atomic sawtooth Au surface can serve as a universal growth template for synthesizing other SC TMdC monolayers, lateral heterostructure, and alloy such as $WSe_2$, $MoS_2$, $MoSe_2/WSe_2$ lateral heterostructure, and $W_{1-x}Mo_xS_2$ alloy. SC $WSe_2$ was grown with a supply of Se precursor at 780–820 °C under growth conditions similar to that used to produce SC $WS_2$ films (see Experimental Section and **Figure 4**a). SC $MoS_2$ was grown by supplying $H_2S$ gas after spin-coating a liquid Mo precursor, identified by Raman spectra (see Experimental Section and Figures 4b, c). ADF-STEM and SAED analysis of coherent $WSe_2$ grains confirm their GB-free regions (Supporting Information, Figure S14).

To strengthen the growth platform, lateral $MoSe_2/WSe_2$ heterostructure was prepared using different growth conditions. Se pellets were vaporized with elevating temperature after spin-coating of liquid Mo precursor (see Experimental Section and Figure 4d). At the early growth stage, SC $MoSe_2$ is nucleated first with coherent triangular grains. At a final growth temperature of 800 °C for 10 min, native tungsten oxides on the W foil are decomposed to sublimate W atoms.[30] The W atoms are energetically preferred to anchor to the edge of the $MoSe_2$ layer and eventually substitute Mo atoms to form $MoSe_2/WSe_2$ heterostructure, while retaining coherent triangular grains (Figure 4d). Consequently, $WSe_2$ domain is gradually formed, as confirmed by the confocal Raman mapping image of $E^1_{2g}$ phonon in the $WSe_2$ region and spatially resolved confocal Raman spectra (Figures 4e, f). $W_{1-x}Mo_xS_2$ alloy can be also grown by supplying $H_2S$ gas at a growth temperature of 800 °C (see Experimental Section and Figure 4g). Mo precursors and vaporized $WO_x$ are mixed together to form alloy and segregate into a serpentine $MoS_2$ structure in a $WS_2$ lattice, identified by distinct intensity differences in the ADF-STEM image and Raman spectra (Figures 4h-j). Moreover, the absence of GB lines between merged grains indicates the single crystallinity of the alloy, confirmed by ADF-STEM and their stacked FFT patterns (Supporting Information, Figure S15).

We propose an atomic sawtooth Au surface as a universal growth platform for SC TMdC monolayers, including $WS_2$, $WSe_2$, $MoS_2$, $MoSe_2/WSe_2$ heterostructure, and $W_{1-x}Mo_xS_2$ alloy in a centimeter scale, regardless of Miller indices. Energetically favorable adsorption sites at periodic gullet step edges promote the growth of coherent TMdC triangular grains to eventually form the GB-free SC TMdC film. Avenues to explore in the future include i) the control of atomic sawtooth Au surface orientation, ii) the scaling up of the fabrication process for industrial applications, and iii) the discovery of appropriate atomic sawtooth surfaces beyond Au films. Our strategy offers new insights into the synthesis of SC diatomic 2D materials in a wafer scale to further facilitate the realization of SC vdW heterostructures.

**Experimental Section**

*Cleaning procedure of Au and W foils:* To remove impurities on the surfaces of Au and W foils, $1.2 \times 1.2$ cm$^2$ high-purity Au foils (0.2 mm thick, 99.99 %, iNexus, Inc.) and $1.3 \times 1.3$ cm$^2$ W foils (0.1 mm thick, 99.95%, Alfa Aesar) were respectively dipped into Au etchant (GE-8111, Transene) and Ni etchant (TFB, Transene) for 10 min. After removal of the impurities, the etchant residues were washed away with fresh deionized water. The cleaned Au and W foils were annealed at 1000 °C for 1 h under high-purity Ar and H$_2$ atmosphere with flow rates of 390 and 10 sccm, respectively.

*Preparation of the atomic sawtooth Au surface via solidification of liquid Au:* Au foil stacked on W foil was loaded into the center of a 2-inch quartz tube. The residual gases in the quartz tube were purged using high-purity Ar (99.999 %) with a flow rate of 500 sccm for 30 min. The furnace was ramped to 1080–1100 °C for 1 h and maintained for 20 min. The temperature was then slowly reduced to 1050 °C with a cooling rate of 0.5–60 °C/min. The furnace was naturally cooled down to room temperature from 1050 °C.

*Precursor preparation:* Prior to chemical-vapor-deposition (CVD) growth, transition metal–precursor solutions (sodium molybdate dihydrate (Na$_2$MoO$_4$·2H$_2$O, ≥ 99.5 %, Sigma Aldrich) for Mo and sodium tungstate dihydrate (Na$_2$WO$_4$·2H$_2$O, ≥ 99 %, Sigma Aldrich) for W, with 2 wt% in acetylacetone) were spin-coated onto atomic sawtooth Au substrates with 2500 rpm for 60 sec.[31,32] Ammonium sulfide solution (($NH_4$)$_2$S, 20 wt% in $H_2O$, Sigma Aldrich) and hydrogen sulfide ($H_2S$) were used as sulfur precursors. ($NH_4$)$_2$S was introduced through bubbling system. For a selenium precursor, selenium pellets ($< 5$ mm, $\geq 99.99\%$, Sigma Aldrich) were evaporated at 360 °C in a two-zone furnace system.

*Growth procedures of various single-crystal (SC) transition metal–dichalcogenide (TMdC) monolayers:* Metal precursor–coated Au substrate was loaded into the quartz tube and purged by high purity Ar (99.999 %) with a flow rate of 500 sccm for 15 min. The furnace was ramped to the growth temperatures (750–880 °C) and maintained with a supply of chalcogen precursor during the growth of TMdCs for 5–20 minutes. After growth, furnace was naturally cooled down to the room temperature. The overall procedure was performed under Ar and $H_2$ atmosphere with flow rates of 350–500 and 0–20 sccm, respectively. Coherent $WS_2$ grains were grown on W precursor–coated Au/W substrate by supplying ($NH_4$)$_2$S at a flow rate of 2–5 sccm at 800–880 °C for 5–20 min.[33] To confirm the presence of grain boundaries in polycrystalline $WS_2$ film, the growth temperature was reduced to 750 °C while retaining the other growth conditions. Two-zone furnace was used to evaporate Se in the zone upstream from the substrate.[34] The growth of $WSe_2$ grains on W precursor–coated Au/W substrate was conducted at 780–820 °C for 7–10 min under a supply of Se vapor. Coherent $MoS_2$ grains were grown on Mo precursor–coated Au/W substrate by supplying $H_2S$ gas at a flow rate of 5–20 sccm at 800 °C for 10 min. To eliminate any interference by W source from native tungsten oxides on W foils, the substrate was annealed at 800 °C for 5 min prior to growing monolayer $MoS_2$. Only Ar gas was used during the entire growth procedure. For the growth of $MoSe_2$/$WSe_2$ heterostructure, triangular $MoSe_2$ grains were first grown on Mo precursor–coated Au/W substrate by supplying Se vapors while the temperature was increased to 800 °C in a two-zone furnace. The $MoSe_2$/$WSe_2$ heterostructure was subsequently grown with a supply of W source from W foil underneath the Au foil via sublimation of native tungsten oxides at 800 °C for 10 min.[30] Coherent $W_{1-x}Mo_xS_2$ grains were grown on Mo precursor–coated Au/W substrate by supplying $H_2S$ gas at a flow rate of 5–20 sccm at 800 °C for 10 min under a mixture of Mo precursor and W, derived from the native tungsten oxides of the W foil.

*Transfer of SC TMdC grown on Au:* As-grown SC TMdC on Au/W substrates were coated with poly(methyl methacrylate) (A9 PMMA, MicroChem) under 4000 rpm for 1 min. The PMMA layer was baked at 200 °C for 5 min. For bubble transfer, the PMMA/TMdC/Au/W sample and Pt foil were used as anode and cathode, respectively. Under a voltage of 3–7 V in 1 M NaOH aqueous solution, the PMMA/TMdC was delaminated from Au substrate by the hydrogen bubbles generated at the interface between the TMdC film and Au substrate. The PMMA/TMdC samples were cleaned by deionized water for 10 min and transferred onto target substrates such as TEM grids and $SiO_2$/Si substrates. To reuse the Au substrates, the Au/W substrate was cleaned in a piranha solution for 3 hours, 10 % nitric acid for 1 h, and buffer oxide etchant for 1 h.

*Characterizations:* Field-emission scanning electron microscopy (FESEM, JSM7000F, JEOL) equipped with the electron backscatter diffraction (EBSD) was employed to analyze the surface orientation of atomic sawtooth Au substrates. FESEM (JSM7100F, JEOL) was used to characterize the surface morphology and orientation of TMdC grains. The optical properties of TMdC grains were characterized by Raman spectroscopy system using 532 nm wavelength (NTEGRA Spectra, NT-MDT). Scanning tunneling microscope (STM, VT-STM, Omicron, Germany) and low-energy electron diffraction (LEED, Specs, Germany), with a beam diameter of ~1 mm, were utilized to corroborate the crystallographic relationship between $WS_2$ and atomic sawtooth Au substrates under ultrahigh vacuum conditions with a base-pressure of ~1.0 $\times 10^{-10}$ Torr at room temperature. The LEED map of a $WS_2$ film for $5 \times 5$ mm$^2$ area was obtained by changing the sample position with a precision of 0.01 mm. The crystal lattice and orientation of TMdC films were characterized by STEM (JEM ARM 200F, JEOL) with an electron acceleration voltage of 80 kV. Cross-sectional STEM lamellas of $WS_2$/Au samples were prepared using a focused ion beam (FIB, FEI Helios NanoLab 450). To visualize the grain boundaries in poly-crystal $WS_2$, as-grown $WS_2$ grains were oxidized in air at 350 °C for 30 min.

*Simulation of STM image:* Simulation of STM images was performed by Tersoff–Hamann approximation using Vienna *ab initio* simulation package (VASP) based on density functional theory (DFT) with a plane-wave basis set.[35-39] We used projector augmented wave (PAW) formalism for the pseudopotentials with Perdew–Burke–Ernzerhof (PBE) parametrization of the general gradient approximation (GGA) as implemented in VASP.[40-42] A slab geometry with a vacuum of ~13 Å was applied to the structural model which consists of monolayer $WS_2$ and 6 layers of Au. The supercell size of $WS_2$ on Au substrate corresponds to the doubled period of Au (221) across the step edges. The supercell size was minimized by compressing $WS_2$ along the step edges due to the heavy computational load, which does not violate the conclusion. Bias for simulated STM images was set ~0.2 eV larger than conduction band minimum similar to experimental conditions. The plane-wave cut-off energy was 400 eV. For the surface Brillouin zone integration, a $1 \times 2$ grid in Monkhorst–Pack special *k*-point scheme, including $\Gamma$-point, was used as implemented in VASP. All atoms were fully relaxed until the residual forces of each atom were less than 0.01 eV/Å. Energy convergence was achieved with a tolerance of $10^{-6}$ eV.

*Computational details:* Adsorption of $W_3S_6$ cluster on the various Au facets were studied by DFT calculations using VASP. The revised PBE exchange-correlation functional and PAW method were used. Periodic supercell geometries were constructed to model (100), (110), (111), (113), (1,4,10), (169), (2,10,15), (221), and (233) Au surfaces using $5 \times 5$, $4 \times 3$, $6 \times 6$, $6 \times 3$, $2 \times 1$, $2 \times 1$, $2 \times 1$, $4 \times 2$, and $4 \times 1$ unit cells, respectively, and a vacuum layer lager than 15 Å was added onto each surface to eliminate interlayer interactions. The Brillouin zones were sampled using $3 \times 3 \times 1$ *k*-point mesh, while the electronic states were smeared using the Methfessel–Paxton scheme with a broadening width of 0.1 eV. The electronic wave functions were expanded on a plane-wave basis set with a cutoff energy of 400 eV, and the atomic-relaxation process continued until the Hellmann–Feynman forces acting on the atoms were less than 0.02 eV/Å.

**Supporting Information**

Supporting Information is available from the Wiley Online Library or from the author.

**Acknowledgements**

K. K. K. acknowledges support by Samsung Research Funding & Incubation Center of Samsung Electronics under Project Number SRFC-MA1901-04, the Basic Science Research Program through the National Research Foundation of Korea (NRF) funded by the Ministry of Science, ICT & Future Planning (2018R1A2B2002302 and 2020R1A4A3079710). S. M. K. acknowledges support by the Basic Science Research Program through the National Research Foundation of Korea (NRF) funded by the Ministry of Science, ICT & Future Planning (2020R1A2B5B03002054). Y. -M. K. acknowledges support by the Creative Materials Discovery Program (2015M3D1A1070672) through the National Research Foundation of Korea (NRF). S. H. C., S. B., S. J. Y., Y. -M. K. & Y. H. L. acknowledge support by the Institute for Basic Science (IBS-R011-D1).

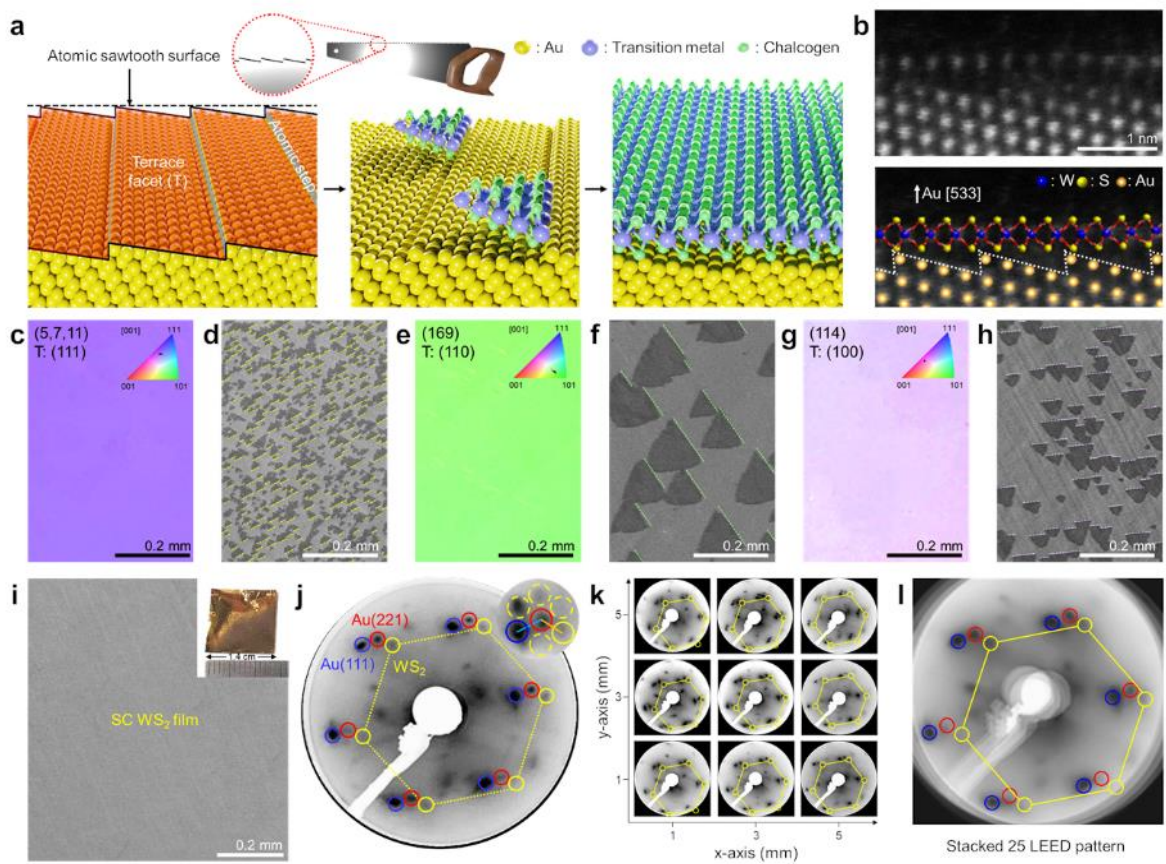

**Figure 1.** Growth of single-crystal $WS_2$ on atomic sawtooth Au surface. a) Schematic illustration of $WS_2$ film grown via an atomic sawtooth Au surface. b) Cross-sectional ADF-STEM image and the corresponding ball-and-stick model of as-grown $WS_2$ on an atomic sawtooth Au (533) surface. c,e,g) EBSD IPF maps and d,f,h) SEM images of coherently aligned $WS_2$ grains on Au (5,7,11), (169), and (114) with respective (111), (110), and (100) terrace facets. i) SEM image of SC $WS_2$ film with a photograph of a centimeter-scale SC $WS_2$ film. j) LEED pattern of $WS_2$ film on Au (221). Each peak is assigned to $WS_2$ hexagonal lattice, Au (111) terrace facet, and atomic sawtooth Au (221) surface. k) Spatial LEED mapping and l) stacked 25 LEED patterns over $5 \times 5$ mm$^2$ area.

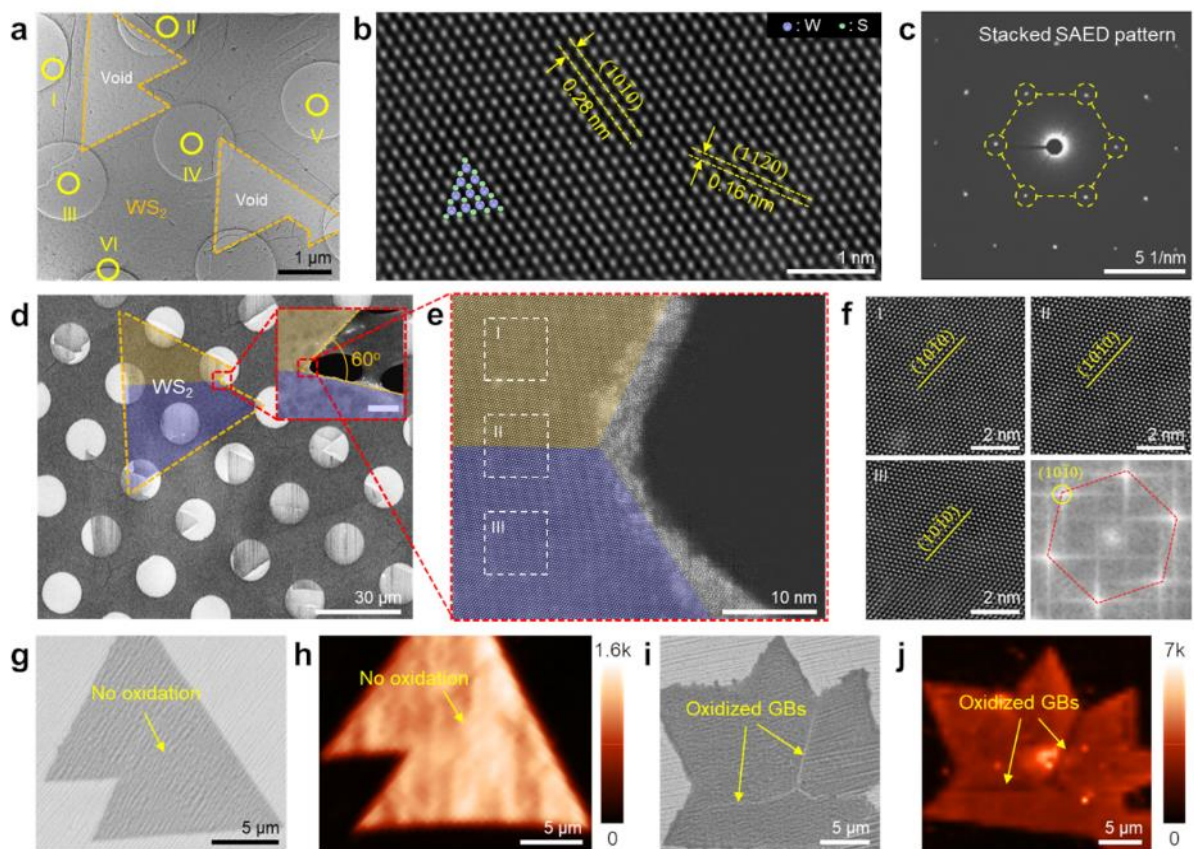

**Figure 2.** Coherent WS$_2$ grains. a) TEM image of aligned WS$_2$ grains transferred onto TEM grid. b) ADF-STEM image of WS$_2$. The d-spacings of the (10-10) and (11-20) planes of WS$_2$ are 0.28 and 0.16 nm, respectively. c) Stacked SAED patterns of six yellow-circled regions in (a). d) SEM image of WS$_2$ grains on TEM grid. Inset shows TEM image at the merged region. The angle between two merged WS$_2$ grains is 60°. e) ADF-STEM image corresponding to the red-dashed square in (d). f) Three ADF-STEM images and stacked FFT patterns obtained from areas enclosed in white-dashed squares in (e). g,i, SEM and h,j, confocal Raman intensity mapping associated with the 2LA(M) mode of g,h) aligned and i,j) misaligned WS$_2$ grains after oxidation at 350 °C.

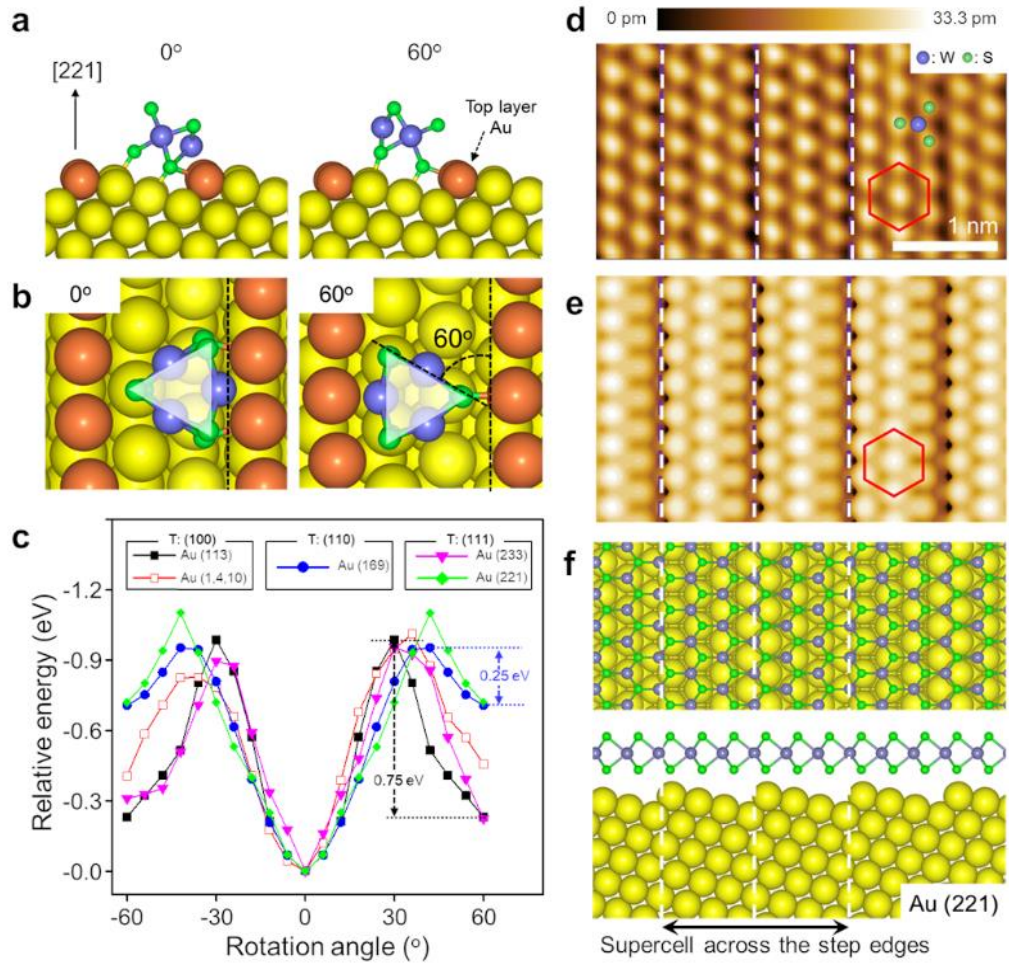

**Figure 3.** Energetics of rotation angles of $WS_2$ clusters at the step edge. a) Side and b) top views of $W_3S_6$ cluster on Au (221) surface with rotation angles of 0° and 60°, respectively. c) Relative binding energies of a $W_3S_6$ cluster on different atomic sawtooth Au surfaces as a function of rotation angle. d) Atomic-resolution STM image of SC $WS_2$ film on Au (221) with applied sample bias of 0.5 V. Bright and dark sites correspond to respective S and W atoms. Au (221) step edges and aligned sulfur atoms are indicated by white-dashed lines and a red hexagon, respectively. e) Simulated STM image and f) corresponding structural model. The coinciding supercell and doubled terrace width of the experimental and simulated results are indicated by dashed lines.

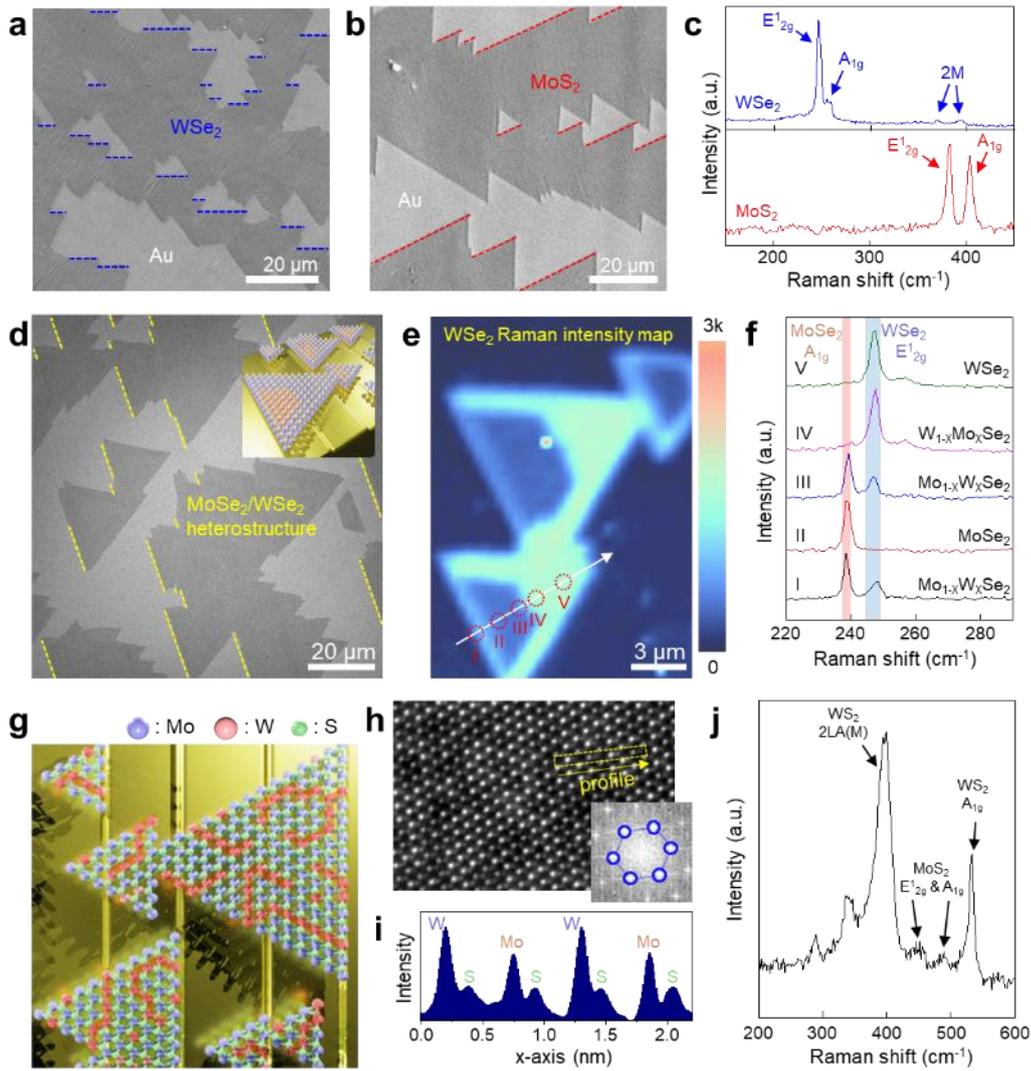

**Figure 4.** Synthesis of SC TMdC monolayers, heterostructure, and alloy on atomic sawtooth Au surfaces. a,b) SEM images and c) Raman spectra of as-grown coherent WSe₂ and MoS₂ grains. d) SEM image and schematic illustration of as-grown MoSe₂/WSe₂ heterostructure. e) Raman mapping image of the WSe₂ E¹₂g phonon of the MoSe₂/WSe₂ heterostructure and f) spatially resolved Raman spectra along white arrow in (e). g) Schematic illustration of serpentine MoS₂ in WS₂ (W₁₋ₓMoₓS₂ alloy). h) ADF-STEM image of W₁₋ₓMoₓS₂ alloy with FFT pattern and i) intensity profile along the yellow-rectangular line in (h). j) Representative Raman spectrum.

# Epitaxial single-crystal growth of transition metal dichalcogenide monolayers *via* atomic sawtooth Au surface

Soo Ho Choi, Hyung-Jin Kim, Bumsub Song, Yong In Kim, Gyeongtak Han, Hayoung Ko, Stephen Boandoh, Ji Hoon Choi, Chang Seok Oh, Hyun Je Cho, Jeong Won Jin, Seok Joon Yun, Bong Gyu Shin, Hu Young Jeong, Young-Min Kim, Young-Kyu Han*, Young Hee Lee*, Soo Min Kim*, and Ki Kang Kim*

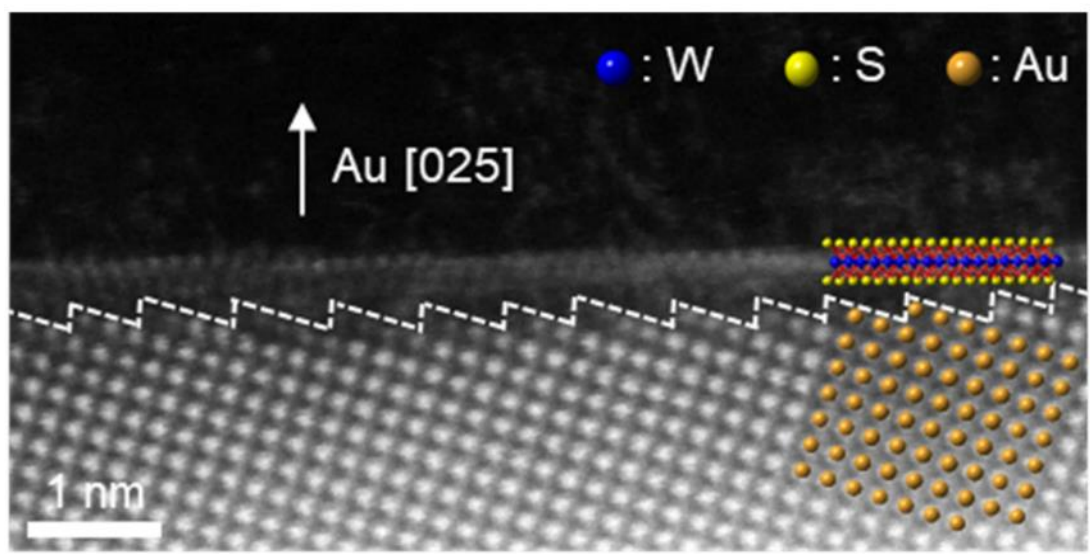

**Figure S1.** Cross-sectional ADF-STEM image of the topmost $WS_2$ on atomic sawtooth Au (025) surface. Atomic sawtooth surface is clearly visualized with periodically spaced gullet-tooth, although the periodicity is slightly irregular, i.e., distances between teeth differ slightly. The topmost $WS_2$ monolayer is fully relaxed on the surface without being corrugated.

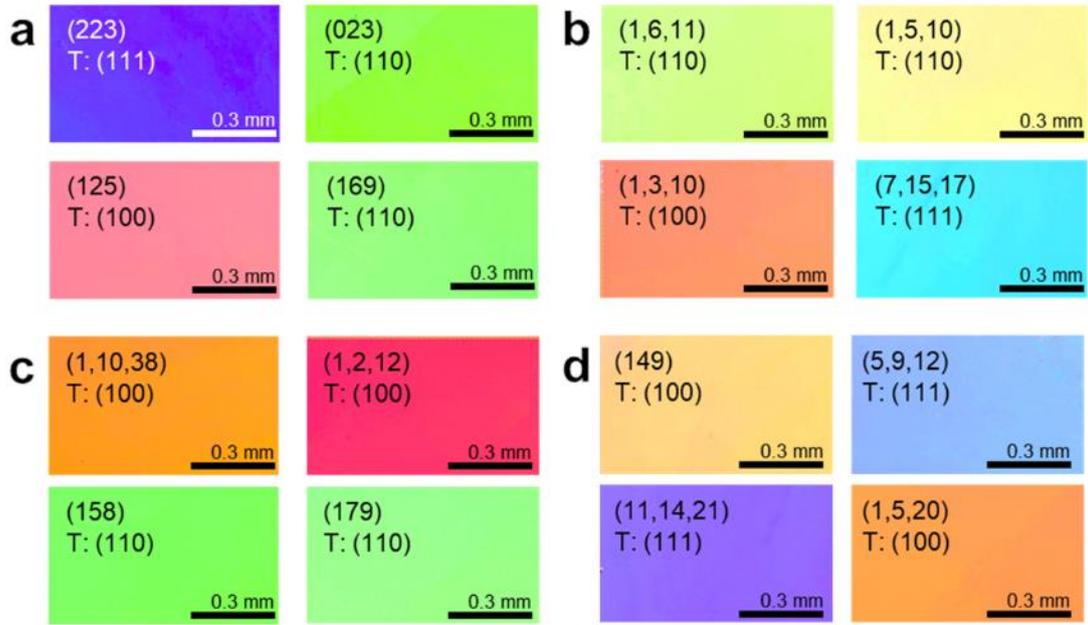

**Figure S2.** The surface orientation of solidified Au under different melting temperatures and cooling rates. a) EBSD IPF maps of solidified Au after melting at 1100 °C with a cooling rate of 60 °C/min. b-d) EBSD IPF maps of solidified Au substrates after melting at 1080 °C with b) fast cooling rate (60 °C/min), c) medium cooling rate (3 °C/min), and d) slow cooling rate (0.5 °C/min). While the terrace facets and Miller indices at the atomic sawtooth surfaces vary with various solidification conditions, they were homogeneous in the entire region that will eventually end up with SC TMdC growth.

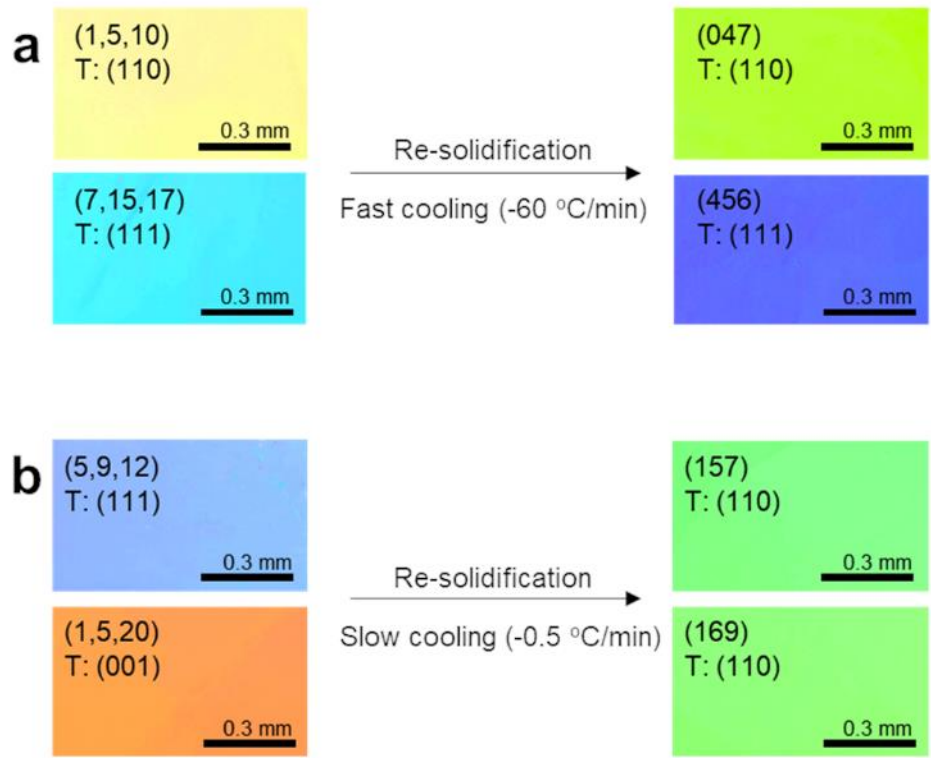

**Figure S3.** Surface orientation of Au after second solidification. a,b) EBSD IPF maps of Au substrates before and after second solidification with a cooling rate of a) 60 °C/min and b) 0.5 °C/min. The surface orientations are not preserved but they are still homogeneous across the entire region.

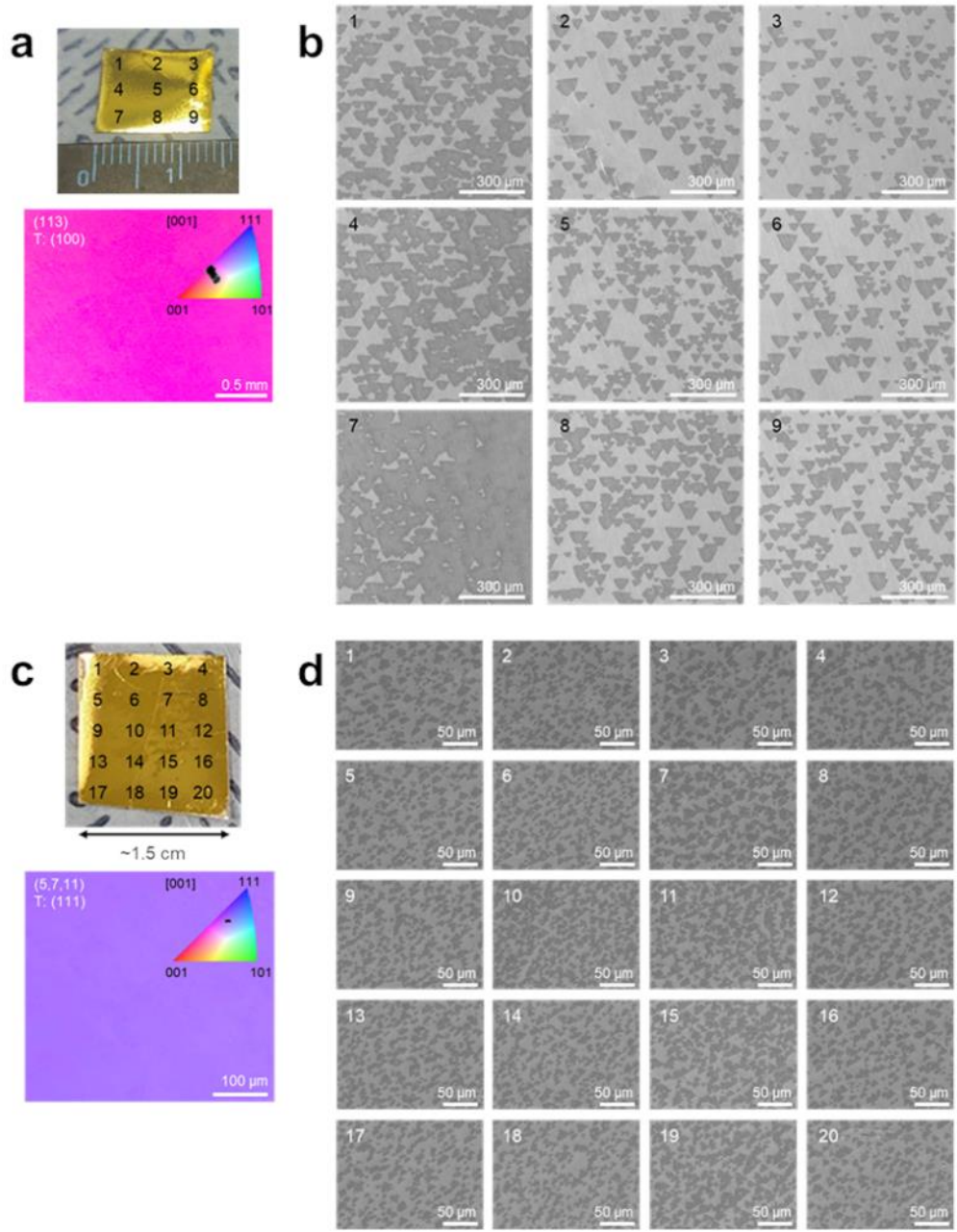

**Figure S4.** Aligned WS$_2$ grains over centimeter-scale atomic sawtooth Au substrates. a,c) Photographs and corresponding EBSD IPF maps of centimeter-scale atomic sawtooth Au foils; a) Au (113) surface and c) Au (5,7,11) surface. b,d) SEM images of WS$_2$ grains grown on respective Au substrates. The WS$_2$ grains on b) Au (113) and d) Au (5,7,11) surfaces are 96.4 % (1709/1772 grains) and 98.8 % (6637/6716 grains) aligned, respectively.

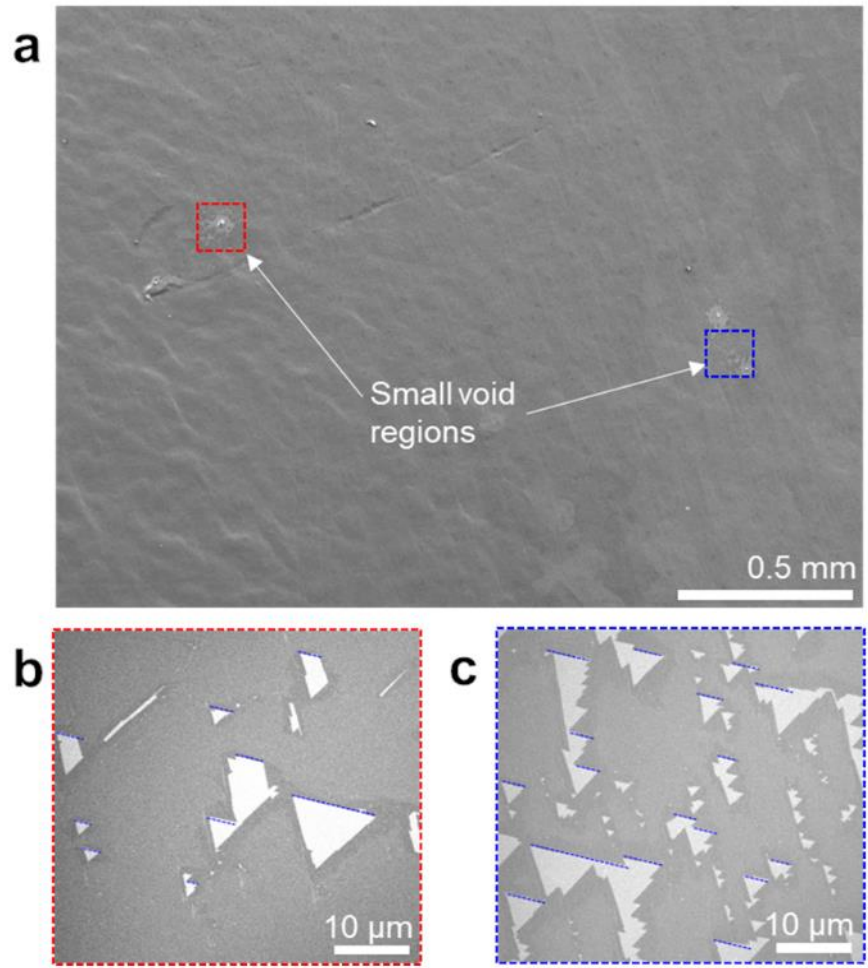

**Figure S5.** Alignment of WS$_2$ grains with void regions prior to the final stage of WS$_2$ film growth. a) Low-magnification SEM image of SC WS$_2$ film. b,c) High-magnification SEM images of void regions corresponding to the areas enclosed in red- and blue-dashed squares in (a). The WS$_2$ grains in the two void regions are still unidirectionally aligned with homogeneous triangular voids, further supporting SC WS$_2$ film growth.

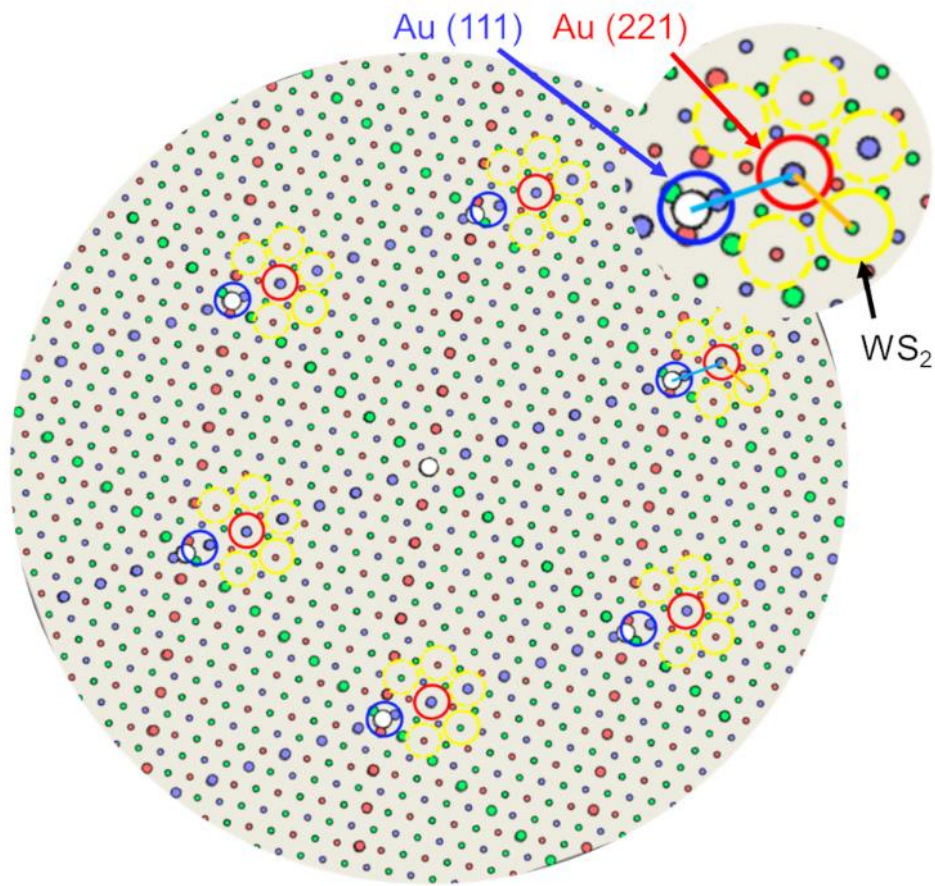

**Figure S6.** LEED analysis of WS$_2$ film. Constructed LEED pattern based on WS$_2$/Au (221). The variation of intensity associated with energy dependence was not considered. The inset shows a supercell consisting of Au (111) (blue), Au (221) (red) and WS$_2$ (yellow). The distance ratio of the orange to blue lines in a supercell is 0.74, in good agreement with LEED pattern presented in Figure 1j.

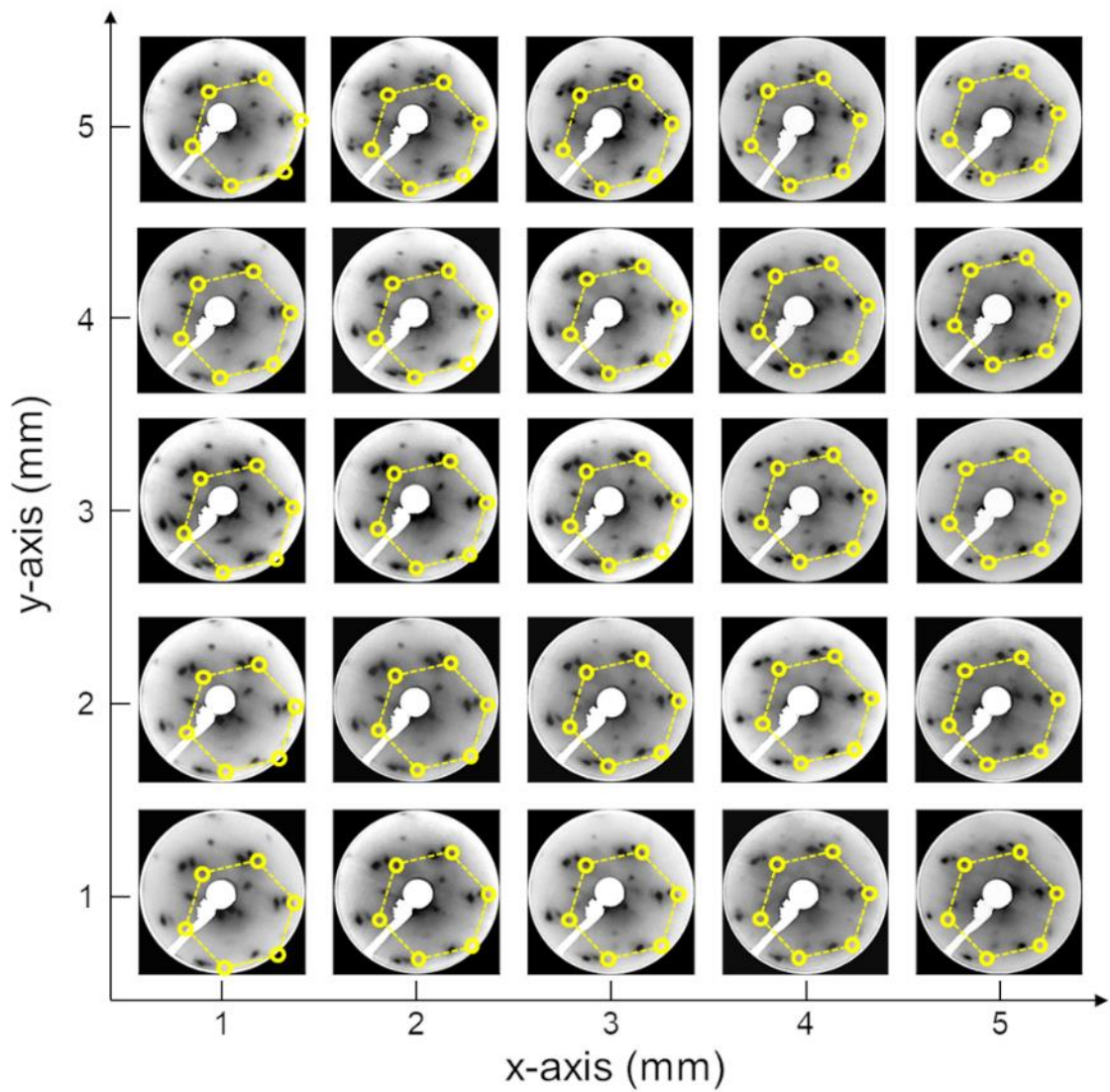

**Figure S7.** Large-area LEED mapping of WS$_2$ film. Spatial LEED mapping images over $5 \times 5$ mm$^2$ area. Coherent hexagonal patterns are clearly observed with 25 superimposed LEED patterns.

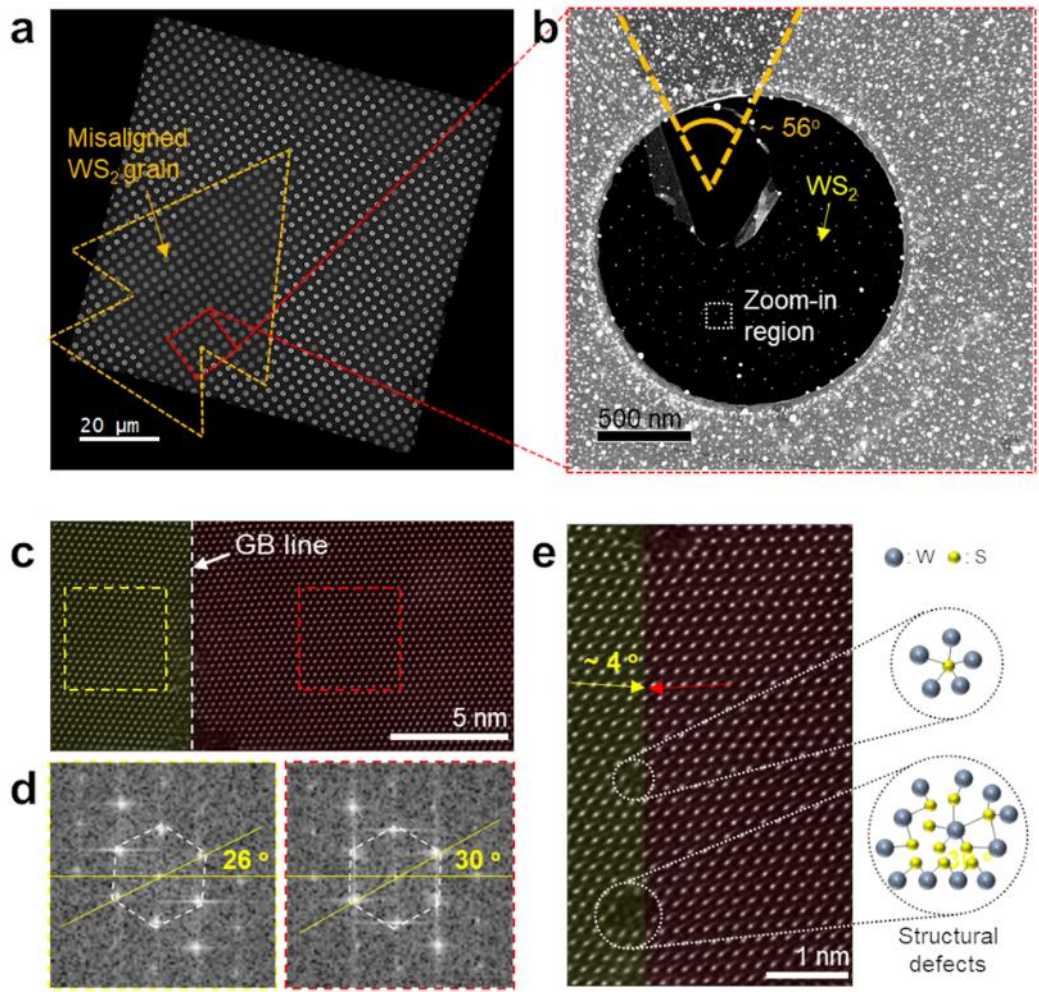

**Figure S8.** Presence of grain boundary line in misaligned WS$_2$ grains. a) Low magnification TEM image and b) ADF-STEM image of misaligned WS$_2$ grains after transfer onto a TEM grid. c) ADF-STEM image at merged region enclosed in a white-dashed square in (b). The GB line is clearly observed. d) FFT patterns extracted from yellow- and red-dashed squares in (c). FFT patterns confirm the misalignment of WS$_2$ grains by ~4°. e) Zoomed-in ADF-STEM image of GB line. Structural defects at GB line are clearly observed.

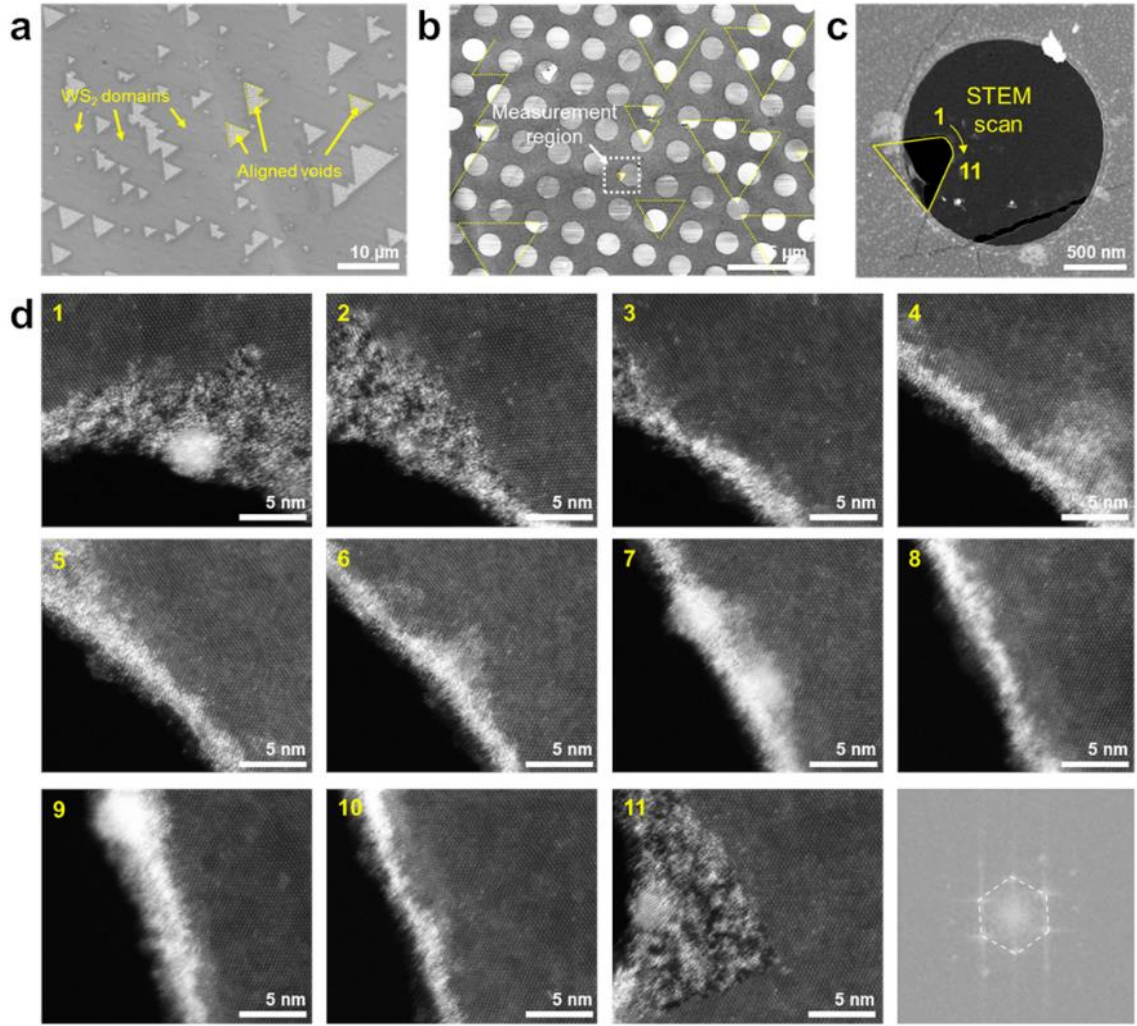

**Figure S9.** STEM analysis of aligned $WS_2$ grains. SEM images of a) as-grown $WS_2$ grains and b) transferred $WS_2$ grains onto a TEM grid. c) Low-magnification ADF-STEM image of $WS_2$ grains at merged region. d) High-magnified consecutive 11 STEM images along the yellow curved arrow and stacked 11 FFT patterns. The stacked FFT pattern shows only one set of hexagonal dots.

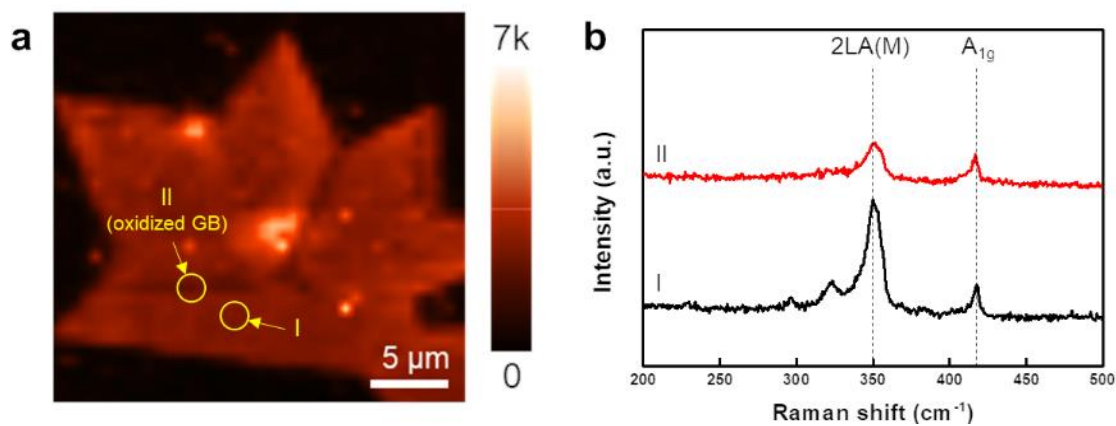

**Figure S10.** Confocal Raman mapping of misaligned $WS_2$ grains after air oxidation. a) Confocal Raman intensity map at 2LA(M) mode of oxidized misaligned $WS_2$ grains. b) Representative Raman spectra of $WS_2$ at the basal plane (I) and oxidized GB line (II). The two characteristic peaks of 2LA(M) and $A_{1g}$ obtained for the oxidized GB line (II) are weaker and broader than those obtained for the basal plane (I).

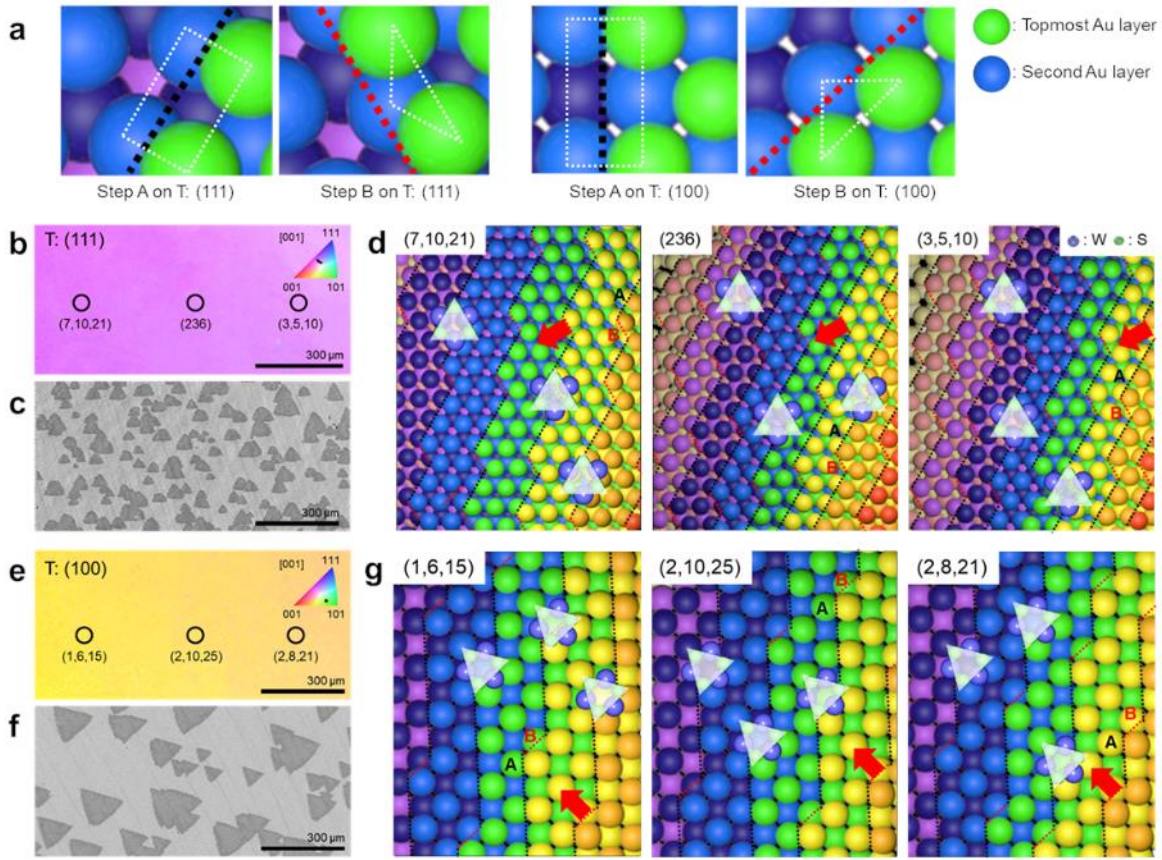

**Figure S11.** Growth of WS$_2$ grains on various Miller indices of a fixed terrace facet. a) Schematic illustration of A and B step edges on fixed (111) and (100) terrace facets. b) EBSD IPF maps and c) corresponding SEM image of coherently aligned WS$_2$ grains on fixed (111) terrace facet grown on Au (7,10,21), (236), and (3,5,10) indices with d) atomic models. e) EBSD IPF map and f) corresponding SEM image of coherently aligned WS$_2$ grains on fixed (100) terrace facet grown on Au (1,6,15), (2,10,25), and (2,8,21) indices with g) atomic models. Regardless of Miller indices, W$_3$S$_6$ clusters are always preferred to anchor to B step edges, based on the binding energy difference of 0.33–0.61 eV from the DFT calculations (Supporting Information, Table S2), indicating that B step edges serve as nucleation sites rather than A step edges. Therefore, coherent WS$_2$ grains are always grown.

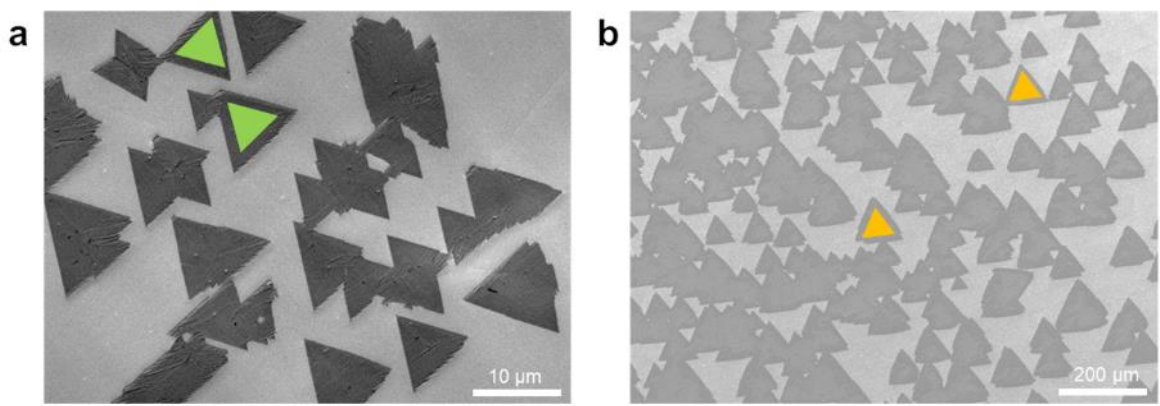

**Figure S12.** Temperature dependence of the growth of WS$_2$. a,b) SEM images of WS$_2$ grains grown at (a) 750 °C and (b) 880 °C, respectively. Energetically stable bimodal angles at 0° and 60° with different energies give rise to inverted triangles at low temperature (a), while at high temperature, unidirectional triangular grains are formed (b).

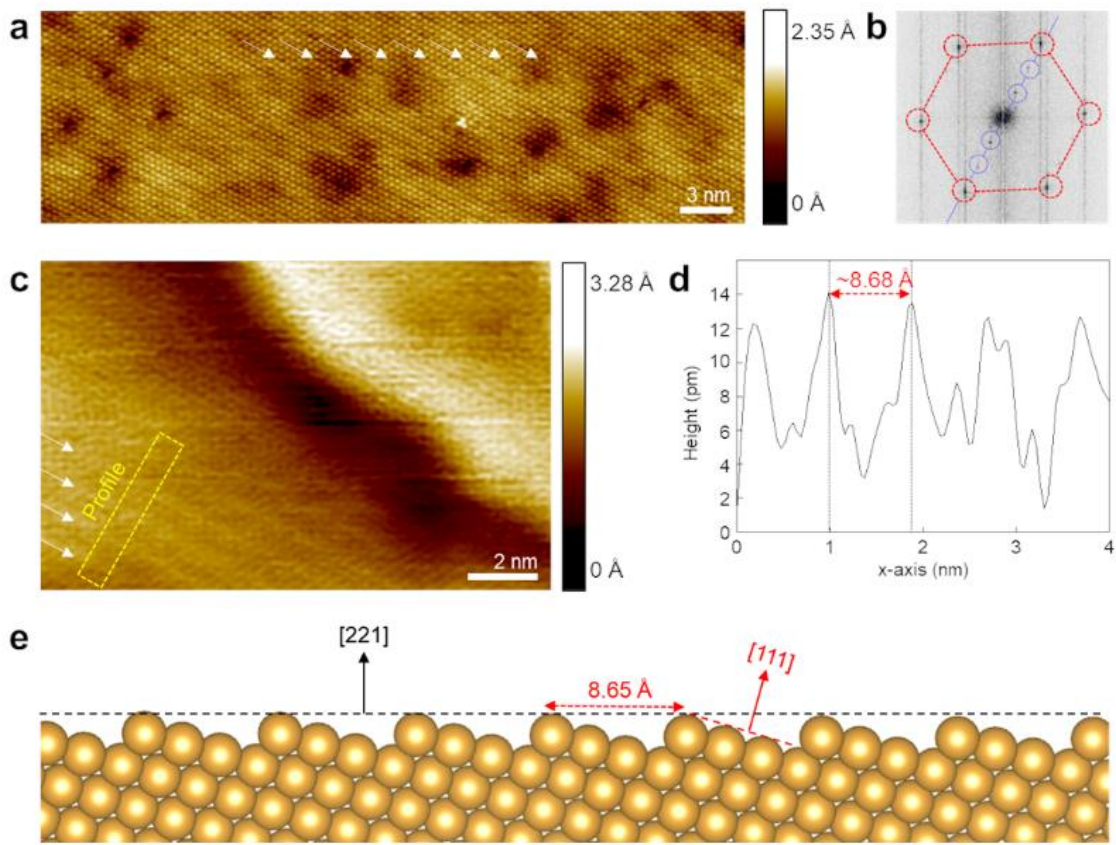

**Figure S13.** STM analysis of epitaxially grown $WS_2$ on Au (221). a) STM image of $WS_2$ grown on Au (221). The substrate-induced stepped edge is indicated by arrows. b) FFT patterns of (a) clearly show × 3 $WS_2$ across the step edges of Au as indicated by blue circles. c) Amplified STM image of $WS_2$ and d) corresponding line profile indicates the substrate-induced modulation of ~8.6 Å with a periodicity of Au (221) along the perpendicular direction to the step edges. e) Side view of the structural model of Au (221).

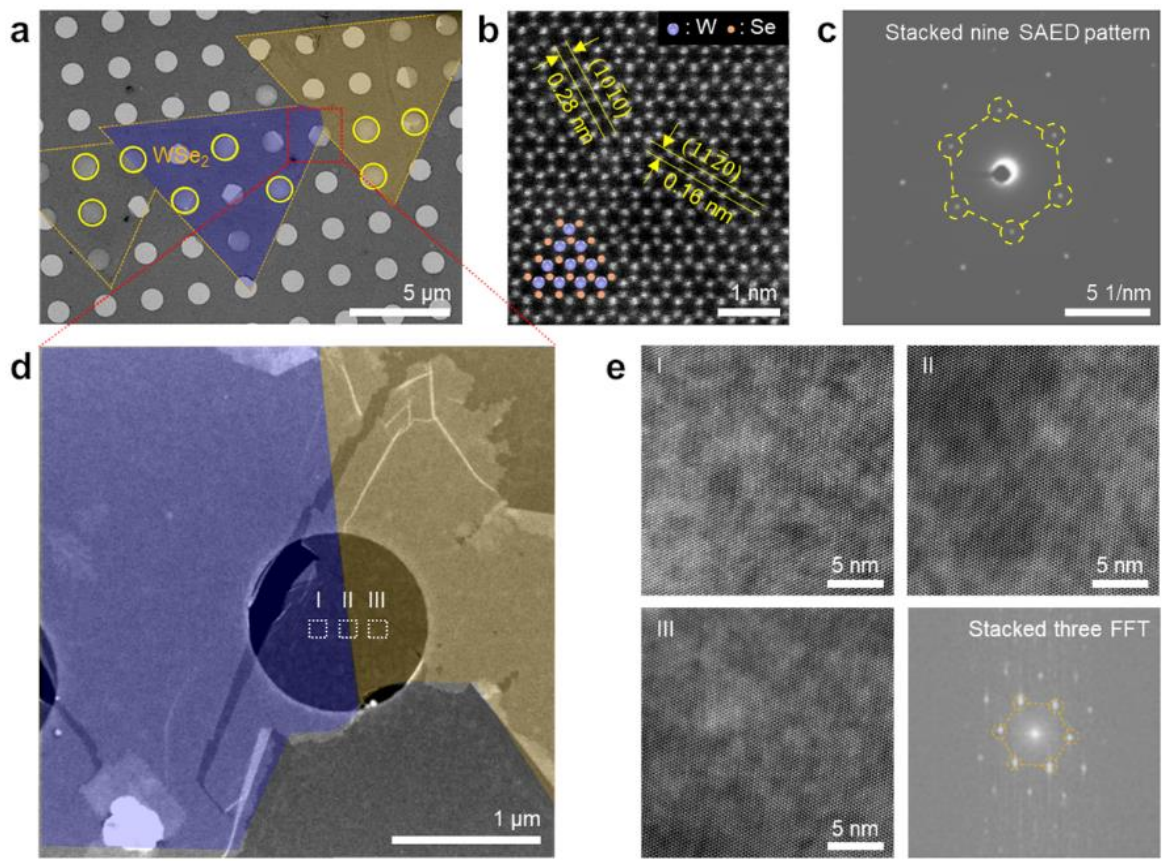

**Figure S14.** Coherent WSe$_2$ grains. a) SEM image of transferred WSe$_2$ grains onto a TEM grid. b) ADF-STEM image of WSe$_2$. c) Stacked SAED pattern of nine yellow-circled regions in (a). d) ADF-STEM image of WSe$_2$ grains at merged region in (a) enclosed in red-dashed square. e) ADF-STEM images of WSe$_2$ from regions I to III in (d) and their stacked FFT image. The presence of noticeable GB lines in the merged region are not found.

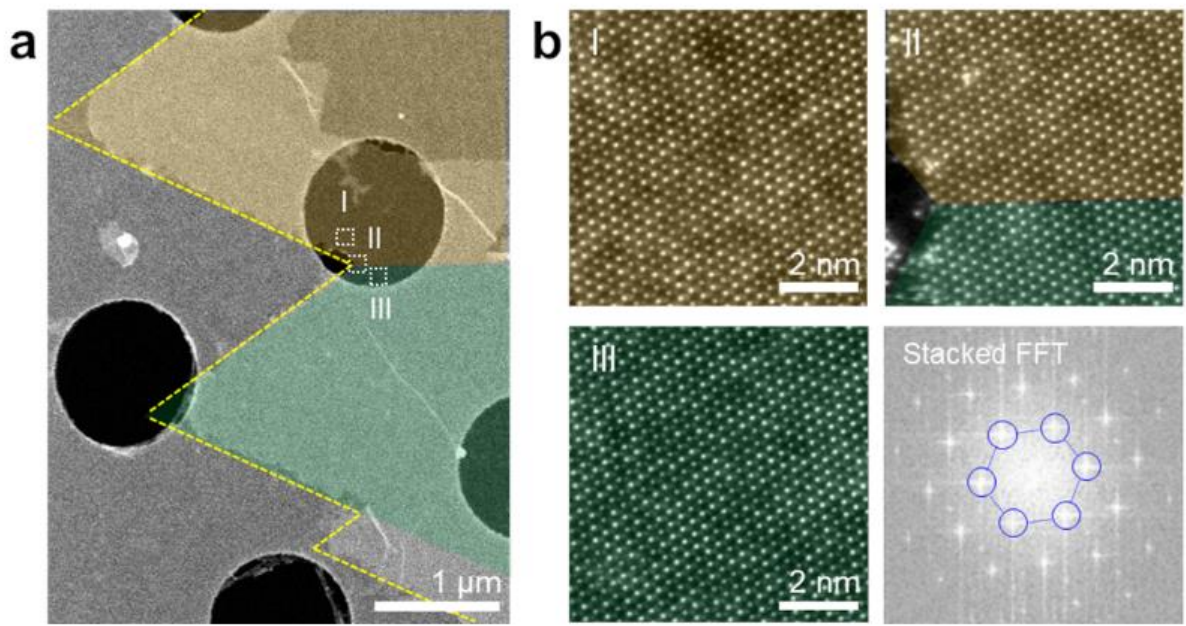

**Figure S15.** Coherent $W_{1-x}Mo_xS_2$ grains. a) Low-magnification ADF-STEM images of transferred $W_xMo_{1-x}S_2$ grains on TEM grid. b) ADF-STEM image of $W_xMo_{1-x}S_2$ from three regions I to III in A and their stacked FFT image. Noticeable GB lines at the merged region are not apparent, indicating that $W_xMo_{1-x}S_2$ grains are merged via seamless-stitching.

**Table S1.** Surface Miller indices of solidified Au with different conditions.

| solidification repetition | Melting temperature | Cooling rate | Miller index |
|---|---|---|---|
| First | 1100 °C | Fast (-60 °C/min) | (223), (023), (125), (169) |
| | 1080 °C | Fast (-60 °C/min) | (1,3,10), (1,6,11), (1,5,10), (7,15,17) |
| | | Medium (-3 °C/min) | (1,10,38), (1,2,12), (158), (1,7,29) |
| | | Slow (-0.5 °C/min) | (11,14,21), (149), (5,9,12), (1,5,20) |
| Second | 1080 °C | Fast (-60 °C/min) | (1,5,10), (7,15,17) → (456), (047) |
| | | Slow (-0.5 °C/min) | (5,9,12), (1,5,20) → (157), (169) |

**Table S2.** Relative total energies of a $W_3S_6$ cluster adsorbed on A and B step edges on atomic sawtooth Au surface including (111), (110), and (100) terrace facets.

| Miller index | Terrace facet | Relative energy (eV) | |
| --- | --- | --- | --- |
| | | A step edge | B step edge |
| (236) | 111 | 0.61 | 0 |
| (169) | 110 | 0.35 | 0 |
| (1,6,15) | 100 | 0.33 | 0 |

**Table S3.** Terrace facets, terrace widths of atomic sawtooth Au surfaces, relative total energies and rotational barrier heights of a $W_3S_6$ cluster adsorbed on sawtooth Au surfaces.

| Miller index | Terrace facet | Effective terrace width (Å) | Relative energy (eV) | | Rotational barrier height (60° → 0°) |
|---|---|---|---|---|---|
| | | | 0° | 60° | |
| (113) | 100 | 2.8 | 0 | 0.23 | 0.75 |
| (1,4,10) | 100 | 4.3 | 0 | 0.22 | 0.73 |
| (169) | 110 | 4.7 | 0 | 0.45 | 0.56 |
| (233) | 111 | 12.1 | 0 | 0.70 | 0.25 |
| (221) | 111 | 7.3 | 0 | 0.73 | 0.38 |